\documentclass[12pt,pdflatex]{iopart}

\usepackage{dcolumn}
\usepackage{bm}
\usepackage{epsfig}
\usepackage{epstopdf}
\usepackage{graphicx}
\usepackage{amssymb}
\usepackage{amsfonts}
\usepackage{hyperref}
\usepackage{color,soul}
\usepackage{appendix}
\usepackage{wrapfig}

\begin{document}
\title{Optimal joint measurements of complementary observables by a single trapped ion}
\author{T. P. Xiong$^{1,2}$, L. L. Yan$^{1,2}$, Z. H. Ma$^{3}$, F. Zhou$^{1}$, L. Chen$^{1}$, W. L. Yang$^{1}$, M. Feng$^{1,4}$ and P. Busch$^{5}$}

\address{$^{1}$ State Key Laboratory of Magnetic Resonance and Atomic and Molecular Physics,
Wuhan Institute of Physics and Mathematics, Chinese Academy of Sciences, Wuhan, 430071, China\\
$^{2}$ University of the Chinese Academy of Sciences, Beijing 100049, China \\
$^{3}$ Department of Mathematics, Shanghai Jiaotong University, Shanghai, 200240, China \\
$^{4}$ Synergetic Innovation Center for Quantum Effects and Applications (SICQEA), Hunan Normal University, Changsha 410081, China \\
$^{5}$ Department of Mathematics and York Centre for Quantum Technologies, University of York, York, United Kingdom }
\ead{zhoufei@wipm.ac.cn}
\ead{mangfeng@wipm.ac.cn}
\ead{paul.busch@york.ac.uk}

\begin{abstract}
The uncertainty relations, pioneered by Werner Heisenberg nearly 90 years ago, set a fundamental limitation
on the joint measurability of complementary observables. This limitation has long been a subject of debate, which has been reignited
recently due to new proposed forms of measurement uncertainty relations. The present work is associated with a new error trade-off relation for compatible observables approximating two incompatible observables, in keeping with the spirit of Heisenberg's
original ideas of 1927. We report the first \textsl{direct} test and confirmation of the tight bounds prescribed by such an error trade-off relation, based on an experimental realisation of optimal joint measurements of complementary observables using a single ultracold $^{40}Ca^{+}$ ion trapped in a harmonic potential. Our work provides a prototypical determination of ultimate joint measurement error bounds with potential applications in quantum information science for high-precision measurement and information security.
\end{abstract}
\maketitle

\section*{1. Introduction}

Quantum measurement, whilst being fundamental to quantum physics, poses perhaps the most difficult problems for the understanding of quantum theory. There are still open questions regarding quantum measurement, such as: why and how does the wave-function collapse happen;  and what exactly are the fundamental precision limits imposed on measurements by the principles of quantum mechanics? With today's rapid progress in technology, high-precision measurements are approaching the ultimate quantum limits. Recent advances, particularly in the area of quantum information science, have led to heightened interest in the fundamental limitations on the achievable quantum measurement accuracy. The unique characteristics of the quantum world, such as Bell non-locality, Einstein-Podolsky-Rosen steering, and entanglement \cite{Gu,Ho}, are actually linked with  uncertainty relations for errors in joint measurements \cite{Wolf,Gu2} more than with the traditional defined uncertainty relations that only address the necessary dispersion in the system observables prior to the measurement. Therefore, scrutinising the lowest error bounds allowed by measurement inaccuracy has become important for current investigations into fundamental quantum physics.

The question of error bounds for joint measurements of incompatible observables was already raised by Werner Heisenberg in 1927, who proposed an answer with his famous uncertainty relation \cite{zphys43-172}. The standard textbook version of the uncertainty relation is  the Robertson-Schr{\"o}dinger inequality: $\Delta\mathcal{A}\Delta\mathcal{B}\ge|\langle [\mathcal{A}, \mathcal{B}]\rangle|/2$, where $\Delta\mathcal{A}$ and $\Delta\mathcal{B}$ are the standard deviations of two non-commuting operators $\mathcal{A}$ and $\mathcal{B}$ and the lower bound is given by the expectation value of the commutator of these operators. This relation concerns separate measurements of $\mathcal{A}$ and $\mathcal{B}$ performed on two ensembles of identically prepared quantum systems. Importantly, it is  conceptually different from Heisenberg's  idea of a trade-off for the errors of approximate simultaneous or successive measurements performed on the same system \cite{pr34-163,zphys44-326,book1928,RMP42-358}.

There are thus two distinct operational aspects of the uncertainty principle \cite{Busch07}:
(a) the preparation of a state with both $\mathcal{A}$ and $\mathcal{B}$ having well defined values is impossible; (b) a measurement of $\mathcal{A}$ inevitably disturbs any subsequent or simultaneous measurement of $\mathcal{B}$. Statement (a) paraphrases the content of the Robertson-Schr\"odinger inequality as an expression of preparation uncertainty. In contrast, (b) points to a necessary trade-off between the inaccuracy in an approximate measurement of $\mathcal{A}$ and the disturbance of a subsequent or simultaneous measurement of $\mathcal{B}$ \cite{prl111-160405}. The latter trade-off constitutes a measurement uncertainty relation (MUR) (or, in the case of successive measurements, more specifically called an error-disturbance relation (EDR)).
\begin{figure}[t]
\centering {\includegraphics[width=8.0 cm, height=5.5 cm]{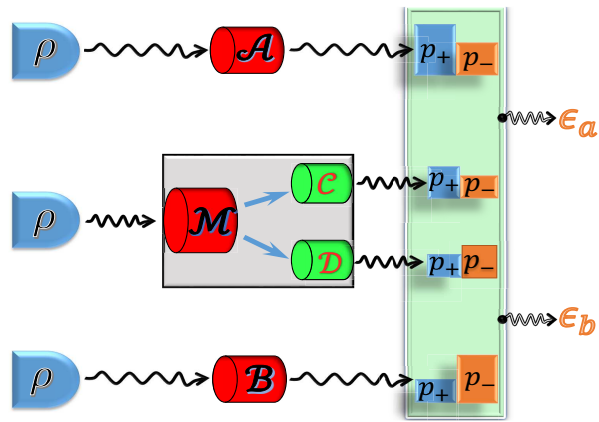}}
\caption{Schematic of optimal joint measurement for verifying a Heisenberg-type measurement uncertainty relation. The quantum apparatus carries out approximate measurements of the incompatible observables $\mathcal{A}$ and $\mathcal{B}$ by joint measurements of the compatible observables $\mathcal{C}$ and $\mathcal{D}$. In our experiment, $\mathcal{C}$ and $\mathcal{D}$ cannot be detected directly, but obtained from the POVM  $\mathcal{M}$ which is measured experimentally. The aim is to obtain optimal joint measurements by choosing appropriate measurement settings so as to minimise the errors $\varepsilon_a$ and $\varepsilon_b$, as defined in the text via the Wasserstein distances employed in the BLW approach.}
\label{fig1}
\end{figure}

In recent years there have been debates over the formulation of  MURs or EDRs due to
disagreements on what constitutes an appropriate quantification of error and disturbance in quantum measurements.
New inequalities for uncertainty relations in the spirit of (b) were independently proposed \cite{prl111-160405,pra67-042105,pla320-367, annphys311-350,pra69-052113,njp12-093011,pra85-062117,pnas110-6742,pra89-012129} with some of them later verified experimentally \cite{natphys8-185,prl109-100404,pra88-022110,sr3-2221,prl110-220402,prl112-020401,prl112-020402,prl116-160405,sa-2016}.
Here we focus on the approach of Busch, Lahti and Werner (BLW) \cite{pra89-012129,rmp86-1261}, which is based on the choice of two compatible observables $\mathcal{C}$ and $\mathcal{D}$ to approximate
two incompatible observables $\mathcal{A}$ and $\mathcal{B}$ (Fig.~\ref{fig1}). The error for (say) $\mathcal{C}$ as an approximation of $\mathcal{A}$ is defined as the worst-case deviation between the probability distributions of $\mathcal{A}$ and $\mathcal{C}$ across all states. Being state-independent quantities, these errors are figures of merit characterising the performance of the measuring device. However, a question arises: how to exactly determine the precise boundary line for the admissible error region \cite{BuHe08,arxiv1,review-busch}, which is crucial not only to the foundation of the Heisenberg's uncertainty relation, but also to realistic quantum operations.

The present work reports the first experimental confirmation of the optimal error bound, based on the realisation of a joint measurement scheme (Fig.~\ref{fig1}) proposed by Yu and Oh \cite{arxiv1}. The Yu-Oh scheme is designed to achieve ultimate lower bounds for the error pairs by optimising the joint measurement. Although the scheme was later clarified with physical interpretation \cite{review-busch}, it remained unclear whether it is adapted to operations in real physical systems. Our experiment utilises the spin of a pure quantum system, a single trapped $^{40}Ca^{+}$ ion. Compared with other experimental setups that operate with ensembles, the single trapped ion can provide more direct and credible evidence as a verification of quantum foundational predictions. By unitary operations under carrier transitions, we demonstrate with high-level control the optimal error bounds achievable by joint measurements of two compatible observables of a qubit encoded in the ion. As witnessed below, our results test precisely the tight bounds of the error trade-off relation and characterise completely the admissible error region. Our experiment constitutes a {\it direct} test of the relevant uncertainty relation in the precise sense explained in \cite{BuSte15}: it provides a comparison of the relevant statistics and hence is based on a true error analysis. In contrast, the tests reported in  \cite{natphys8-185,prl109-100404,pra88-022110,sr3-2221,prl110-220402, prl112-020401,prl112-020402,prl116-160405} regard the purported error quantities as some quantum mechanical expectation values that are to be determined by statistics of experiments that have nothing to do with an error analysis whatsoever \cite{rmp86-1261}. Hence our results are more directly relevant to the exploration of the fundamental
quantum limits of high precision measurements.

\section*{2. System and scheme}
\subsection*{2.1 The experimental system}
\begin{figure*}[tbph]
\centering {\includegraphics[width=12 cm, height=6.5 cm]{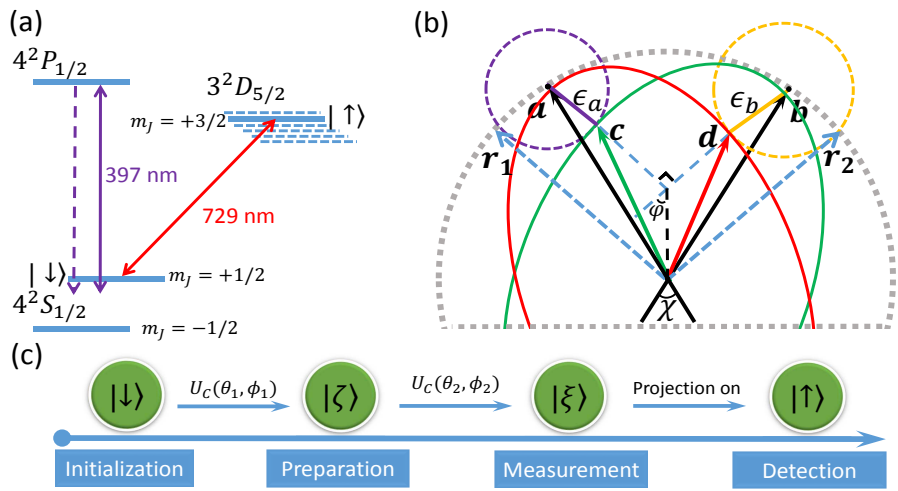}}
\caption{The experimental scheme for verifying the Yu-Oh proposal of optimal joint measurement. (a) Relevant energy levels of the $ ^{40}Ca^+$ ion and transitions. The qubit is encoded in $|4 ^{2}S_{1/2}, m_J= +1/2\rangle=|\downarrow\rangle$ and $|3 ^{2}D_{5/2}, m_J= +3/2\rangle=|\uparrow\rangle$, respectively. A narrow-linewidth 729-nm laser couples the two encoded states under carrier transitions. We measure the probability of the ion in the $|\downarrow\rangle$ state by detecting light spontaneous decay from the level of $4^2P_{1/2}$. (b) The Bloch vectors $\bm{a}, \bm{b}, \bm{c}$ and $\bm{d}$ correspond to the observables $\mathcal{A},\mathcal{B},\mathcal{C}$ and $\mathcal{D}$, respectively. $\chi$ is the angle between $\bm{a}$ and $\bm{b}$, and $\varphi$ is an angle defined in the text. The ellipses enclose the regions in which the vectors $\bm{c}$ and $\bm{d}$ may lie, respectively, according to the compatibility condition $f(\bm{c},\bm{d})=2$, given a fixed value of $\bm{d}$ (the green ellipse) or $\bm{c}$ (the red ellipse).
The errors $\epsilon_a$ and $\epsilon_b$ are given, respectively, by $\epsilon_a=\|\bm{a}-\bm{c}\|$ and $\epsilon_b=\|\bm{b}-\bm{d}\|$, whose operational meaning is explained in the text. (c) Experimental implementation steps and the corresponding states of the system. The ion is first laser-cooled down to nearly the ground state of the quantised vibration. The system starts from the qubit state $|\downarrow\rangle$ and evolves to $|\zeta\rangle$ under the preparation pulse $U_C(\theta_1,\phi_1)$. Then the measurement pulse $U_C(\theta_2,\phi_2)$ rotates the system to the measurement state $|\xi\rangle$, followed by the detection on $|\uparrow\rangle$. See details in Subsection 3.1.}
\label{fig2}
\end{figure*}

A single $^{40}Ca^{+}$ ion is confined in a linear Paul trap, whose axial and radial frequencies are $\omega_z/2\pi=1.01$ MHz and $\omega_r/2\pi=1.2$ MHz, respectively. Under the magnetic field of 6 Gauss, we encode the qubit in $|4 ^{2}S_{1/2}, m_{J}=+1/2\rangle$ as $|\downarrow\rangle$ and in $|3 ^{2}D_{5/2}, m_{J}=+3/2\rangle$ as $|\uparrow\rangle$, where $m_{J}$ is magnetic quantum number (see Fig.~\ref{fig2}(a)). We couple the qubit by a narrow-linewidth 729-nm laser with wave vector $\bm{k}$ at an angle of $22.5^{o}$ to the trap z-axis, resulting in a Lamb-Dicke parameter of $\eta_{z}\approx$ 0.09. After Doppler cooling and optical pumping, the z-axis motional mode is cooled down to the vibrational ground state with the final average phonon number $\bar{n}_{z}<$ 0.1 by the resolved sideband cooling. The ion is initialised to $|\downarrow\rangle$ with a probability about 98.7\%. With the 729-nm laser pulses, the system evolves under the government of the carrier-transition operator
\begin{equation}
U_{C}(\theta,\phi) = \cos(\theta/2)I - i\sin(\theta/2)(\sigma_{x}\cos\phi -\sigma_{y}\sin\phi),
\label{4}
\end{equation}
where $\theta=\Omega t$ is determined by the evolution time, $\Omega$ and $\phi$ are the Rabi frequency representing the laser-ion coupling strength and the laser phase, respectively. Each experimental cycle is synchronised with the 50-Hz AC power line and repeated  40,000 times. The 729-nm laser beam is controlled by a double pass acousto-optic modulator.
The frequency sources for the acousto-optic modulator are based on a direct digital synthesiser controlled by a field programable gate array. Employment of the direct digital synthesiser helps the phase- and frequency-control of the 729-nm laser during each
experimental operation. \\

\subsection*{2.2 The optimal error trade-off relation}
Before presenting our experimental results, we first describe the Yu-Oh proposal briefly in a geometric way, the main idea of which is graphically sketched in Fig.~\ref{fig2}(b). The aim is to find the optimal trade-off between the errors in joint measurement of two incompatible observables. We consider two generally incompatible qubit observables $\mathcal{A}$ and $\mathcal{B}$, represented by the operators $\bm{a}\cdot\bm{\sigma}$ and $\bm{b}\cdot\bm{\sigma}$, respectively. Here $\bm{a}, \bm{b}$ are unit vectors and $\bm{\sigma}=(\sigma_x,\sigma_y,\sigma_z)$ is the vector whose components are the Pauli matrices. The angle between $\bm{a}$ and $\bm{b}$ is specified  by $\sin\chi=\Vert\bm{a}\times \bm{b}\Vert$. As approximations of $\mathcal{A}$ and $\mathcal{B}$, we employ positive operator-valued measures (POVMs) $\mathcal{C}$ and $\mathcal{D}$, whose first moment operators are $\bm{c}\cdot\bm{\sigma}$ and $\bm{d}\cdot\bm{\sigma}$, respectively, with $\Vert \bm{c}\Vert, \Vert\bm{d}\Vert\leq 1$. Thus, as indicated in Fig.~\ref{fig2}(b), the  errors for these observables are given by $\epsilon_a=\Vert\bm{a}-\bm{c}\Vert$ and $\epsilon_b=\Vert\bm{b}-\bm{d}\Vert$, for which the optimal error trade-off relation is given below.

To describe the optimisation procedure, we first recall that the constraint of the compatibility (or joint measurability) of the observables $\mathcal{C}$ and $\mathcal{D}$ is equivalent to the inequality $f(\bm{c},\bm{d})\equiv \Vert\bm{c}+\bm{d}\Vert+\Vert\bm{c}-\bm{d}\Vert\leq 2$.
For any fixed $\bm{c}$ ($\bm{d}$), this inequality defines an ellipsoid
that restricts the possible choices of  $\bm{d}$ ($\bm{c}$). This ellipsoid, centred at the origin, lies inside the unit ball and has the major axis given by a diameter in the direction of $\bm{c}$ ($\bm{d}$). It indicates that optimal approximation errors will be achieved by choosing $\bm{c}$ and $\bm{d}$ in the plane spanned by $\bm{a}$ and $\bm{b}$. Thus one may consider two ellipses in that plane to characterise the compatibility of $\mathcal{C}$ and $\mathcal{D}$, see Fig.~\ref{fig2}(b). To minimise the errors in this case, we draw two circles centred at the end points of the vectors $\bm{a}$ and $\bm{b}$, so that the errors $\epsilon_{a}$ and $\epsilon_{b}$ are the radii. To carry out the optimisation, we fix an error value, e.g., $\epsilon_a$, which can be realised by many values of $\bm{c}$. For each of these possibilities, we can find a specific $\bm{d}$ that gives the smallest value of $\epsilon_b$. Among all those $\epsilon_b$, we choose the smallest. The chosen pair ($\epsilon_a,\epsilon_b$) satisfies the geometric condition, shown in Fig.~2(b), that the two circles are tangent to their corresponding ellipses and $\bm{a}-\bm{c}$ is perpendicular to $\bm{b}-\bm{d}$, and thereby lies on the lower boundary curve of the admissible error region. As such, the optimal, minimal distances from the vectors $\bm{a},\ \bm{b}$  to vectors $\bm{c},\ \bm{d}$ inside the relevant elliptic regions defined by $f(\bm{c},\bm{d})\leq 2$ occur at the boundaries, i.e., $f(\bm{c},\bm{d})=2$. Under this condition, we obtain the optimal results and refer to $(\varepsilon_a,\varepsilon_b)$ as the optimal worst-case (OWC) error pair.

For a state $\rho=(I+\bm{r}\cdot\bm{\sigma})/2$, the error between $\mathcal{A}$ and $\mathcal{C}$ is defined by the Wasserstein 2-distance \cite{review-busch}
\begin{equation}
\Delta_{\rho}(\mathcal{A},\mathcal{C})^2=4|p_{\rho}^{A_+}-p_{\rho}^{C_+}|=2|(\bm{a}-\bm{c})\cdot \bm{r}|.
\label{eq8}
\end{equation}
The quantity $\Delta(\mathcal{A},\mathcal{C})^2$ is the maximum of the above quantity over all  states, which (for $\bm{c}\ne\bm{a}$) actually corresponds to the case of
$\bm{r}$ being parallel to $\bm{a}-\bm{c}$, i.e., $\bm{r}=(\bm{a}-\bm{c})/\Vert\bm{a}-\bm{c}\Vert$. Hence, $\Delta(\mathcal{A},\mathcal{C})^2=2\Vert\bm{a}-\bm{c}\Vert$, and similarly $\Delta(\mathcal{B},\mathcal{D})^2=2\Vert\bm{b}-\bm{d}\Vert$. Here we define the worst-case errors $\epsilon_a=\Delta(\mathcal{A},\mathcal{C})^2/2$ and $\epsilon_b=\Delta(\mathcal{B},\mathcal{D})^2/2$. The approximators $\mathcal{C}$ and $\mathcal{D}$ are generally unsharp observables with the degrees of unsharpness $u_c$ and $u_d$ defined in \cite{review-busch}.

For a given $\varphi$, the OWC errors $(\varepsilon_a,\varepsilon_b)$ are obtained by optimising the vectors $\bm{c}$ and $\bm{d}$ as below \cite{arxiv1},
\begin{eqnarray}
\bm{c}&=\frac{\sin\varphi[\varepsilon_b +(1-h^2)\cos\varphi]\bm{a}+h\varepsilon_a \cos\varphi\bm{b}}{\sin\chi},  \nonumber \\
\bm{d}&=\frac{\cos\varphi[\varepsilon_a +(1-h^2)\sin\varphi]\bm{b}+h\varepsilon_b \sin\varphi\bm{a}}{\sin\chi},  \nonumber
\label{eq9}
\end{eqnarray}
where $h=\cos\chi/\sqrt{1+\sin\chi\sin 2\varphi}$ and the OWC errors $\varepsilon_{a}$ and $\varepsilon_{b}$ are given by
\begin{eqnarray}
\varepsilon_a &=\frac{\sin\varphi+\sin\chi\cos\varphi}{\sqrt{1+\sin\chi\sin2\varphi}}-\sin\varphi, \nonumber \\
\varepsilon_b &=\frac{\cos\varphi+\sin\chi\sin\varphi}{\sqrt{1+\sin\chi\sin2\varphi}}-\cos\varphi.
\label{eq10}
\end{eqnarray}
As proven in \cite{arxiv1}, the OWC error pairs provide an ultimately tight lower bound for the error trade-off relation. The mapping
$\varphi\mapsto (\varepsilon_a(\varphi),\varepsilon_b(\varphi))$ describes a curve in the $(\epsilon_a,\epsilon_b)$-space that marks the boundary between the regions of admissible and inadmissible error pairs for the compatible observables $\mathcal{C}$ and $\mathcal{D}$.

For our purpose, we define $\varphi$ to be the angle that satisfies
$\sin\varphi=\sqrt{(1-\|\bm{d}\|^2)/(1-(\bm{c}\cdot\bm{d})^2)}$ and $\cos\varphi=\sqrt{(1-\|\bm{c}\|^2)/(1-(\bm{c}\cdot\bm{d})^2)}$
for $\bm{c}$ and $\bm{d}$ such that $f(\bm{c},\bm{d})=2$ is given with $\varphi\in[0,\pi/2]$.
Then a Heisenberg-type MUR for the pair of observables can be written as a family of error trade-off relations \cite{arxiv1},
\begin{equation}
\epsilon_a\sin\varphi+\epsilon_b\cos\varphi\geq \sqrt{1+\sin\chi\sin2\varphi}-1,
\label{Eq1}
\end{equation}
which collectively describe the admissible error region. The physical interpretation of this inequality is that of an intricate interplay between the incompatibility of $\mathcal{A},\mathcal{B}$ and the unsharpness of $\mathcal{C},\mathcal{D}$ (required by their compatibility) resulting in a lower bound to the approximation errors $\epsilon_a$ and $\epsilon_b$.
For any value of $\varphi$, the equality in Eq. (\ref{Eq1}) is achieved only for a particular pair ($\varepsilon_{a},\varepsilon_{b}$) of OWC error values. Eq. (\ref{Eq1}) represents a trade-off relation between the errors, i.e., the incompatibility of the target observables $\mathcal{A},\ \mathcal{B}$ (through the term $\sin\chi$ which is a function of the commutator of $\bm{a}\cdot\bm{\sigma}$ and $\bm{b}\cdot\bm{\sigma}$), and the parameter $\varphi$. An interpretation of $\varphi$ can be given in terms of the  degrees of unsharpness of the compatible observables $\mathcal{C}$ and $\mathcal{D}$ (defined by $u_c^2=1-\|\bm{c}\|^2$, $u_d^2=1-\|\bm{d}\|^2$) by observing that $\sin\varphi=u_d/\sqrt{u_c^2+u_d^2}$, $\cos\varphi=u_c/\sqrt{u_c^2+u_d^2}$ \cite{review-busch}.

Putting $\varphi=\pi/4$, Eq. (\ref{Eq1}) reduces to the simple inequality considered in \cite{pra89-012129},
\begin{equation}
\epsilon_a+\epsilon_b\geq \sqrt{2}(\sqrt{1+\sin\chi}-1).
\label{Eq2}
\end{equation}
This relation is weaker than Eq. (\ref{Eq1}), and the straight line defined by it in the $(\epsilon_a,\epsilon_b)$-plane touches the convex region defined by Eq. (\ref{Eq1}) exactly in the point where $\epsilon_a=\epsilon_b=(\sqrt{1+\sin\chi}-1)/\sqrt2$.
The maximal lower bound, i.e., $2-\sqrt{2}$, occurs in the case that the vectors $\bm{a}$ and $\bm{b}$ associated, respectively, with $\mathcal{A}$ and $\mathcal{B}$ are perpendicular to each other ($\chi=\pi/2$).

The OWC errors of the experimental measurement are directly determined from a comparison of the statistics of $\mathcal{A}$ ($\mathcal{B}$) with the statistics of $\mathcal{C}$ ($\mathcal{D}$). The lower bound errors are given by
\begin{equation}
\varepsilon_a=2|p^{A_+}_{\rho_1}-p_{\rho_1}^{C_+}|, \quad \varepsilon_b=2|p^{B_+}_{\rho_2}-p_{\rho_2}^{D_+}|,
\label{eq3}
\end{equation}
where the measurement probability distributions are $p^{X_{\pm}}_{\rho_{1,2}}=$Tr$[X_{\pm}\rho_{1,2}]$ with the quantum states determined by $\rho_{1,2}=(I+\bm{r_{1,2}})/2$ and $X_{\pm}=(I\pm\bm{x}\cdot\bm{\sigma})/2$ for $\bm{x}=\bm{a},\bm{b},\bm{c}$ and $\bm{d}$.
As seen from Fig.~\ref{fig2}(b), the OWC errors defined in Eq.~(\ref{eq3}) appear under the conditions that $\bm{r_1}$ ($\bm{r_2}$) is  parallel to $\bm{a}-\bm{c}$ ($\bm{b}-\bm{d}$) and $\bm{r_1}$ is perpendicular to $\bm{r_2}$.

\section*{3. Experimental implementation}
\subsection*{3.1 The single-qubit measurement}
The incompatible observables $\mathcal{A}$ and $\mathcal{B}$ are directly measured in a single qubit, but compatible observables $\mathcal{C}$ and $\mathcal{D}$ are obtained by joint measurements on a POVM $\mathcal{M}=\{M_{\mu\nu}\}$, given by the rank-1 positive operators
\begin{equation}
M_{\mu\nu}=\frac{1}{4}{\bigl[(1+\mu\nu h)I+(\mu\bm{c}+\nu\bm{d})\cdot\bm{\sigma}\bigr]},
\label{eq11}
\end{equation}
with $\mu,\nu=\pm$1. As such, we obtain the marginality relations $C_{\pm}=M_{\pm+}+M_{\pm-}$ and $D_{\pm}=M_{+\pm}+M_{-\pm}$, where $C_{\pm}=(I\pm\bm{c}\cdot\bm{\sigma})/2$, $D_{\pm}=(I\pm\bm{d}\cdot\bm{\sigma})/2$ and we simply write M$_{\mu\nu}$ as M$_{++}$, M$_{+-}$, M$_{-+}$ or M$_{--}$, depending on the values of $\mu$ and $\nu$. Such a POVM as M$_{\mu\nu}$ is often constructed by employing a second, ancillary qubit. But as designed in the Yu-Oh proposal \cite {arxiv1}, $M_{\mu\nu}$ can be realised in a single qubit without any ancilla, at the expense of losing generality, i.e., no possibility to observe the region above the lower bound. Nevertheless, in this way, the most important characteristic---the lower bound, corresponding to optimal approximations, can be observed in a fundamental single qubit \footnote{As clarified below, if two qubits are employed, with one of them as an ancilla, for constructing the POVM M$_{\mu\nu}$, we may experimentally test more curves
as the error trade-off relations \cite{pra89-012129}, where the lower bounds of the Heisenberg's uncertainty relation are just the OWC errors demonstrated in the present paper. In this sense, our single-ion experiment, with more precision in control than the two-ion counterpart, is the best candidate for verifying the boundary line for the admissible error region of the Heisenberg's uncertainty relation.}. Assuming $\bm{d}\neq\pm\bm{c}$, we can rewrite $M_{\mu\nu}$ as $ M_{\mu\nu}=(1+\mu\nu h)(I+\mu \mathcal{S}^{\mu\cdot\nu})/4 $ with the two operators $\mathcal{S}^{\pm}=(\bm{c}\pm\bm{d})\cdot\bm{\sigma}/\Vert\bm{c}\pm\bm{d}\Vert$. Thus we have
$p^{M_{\mu\nu}}_{\rho}=(1+\mu\nu h)p^{S_{\mu}^{\nu}}_{\rho}/2$ with $S^{\nu}_{\mu}=(I+\mu\mathcal{S}^{\mu\cdot\nu})/2 $.
From the marginality relations, we obtain $p_{\rho}^{C_+}=p^{M_{++}}_{\rho}+p^{M_{+-}}_{\rho}$ and $p_{\rho}^{D_+}=p^{M_{++}}_{\rho}+p^{M_{-+}}_{\rho}$,
which, combined with Eq. (\ref{eq3}), yields
\begin{eqnarray}
\varepsilon_a&=2\left|p^{A_+}_{\rho_1}-P_+p^{S_{+}^{+}}_{\rho_1}-P_-p^{S_{+}^{-}}_{\rho_1}\right|, \nonumber \\
 \varepsilon_b&=2\left|p^{B_+}_{\rho_2}-P_+p^{S_{+}^{+}}_{\rho_2}-P_-p^{S_{-}^{+}}_{\rho_2}\right|,
\label{eq12}
\end{eqnarray}
where $P_{\pm}=(1\pm h)/2$.

\begin{figure*}[htbp]
\centering {\includegraphics[width=11.0 cm, height=5.5 cm]{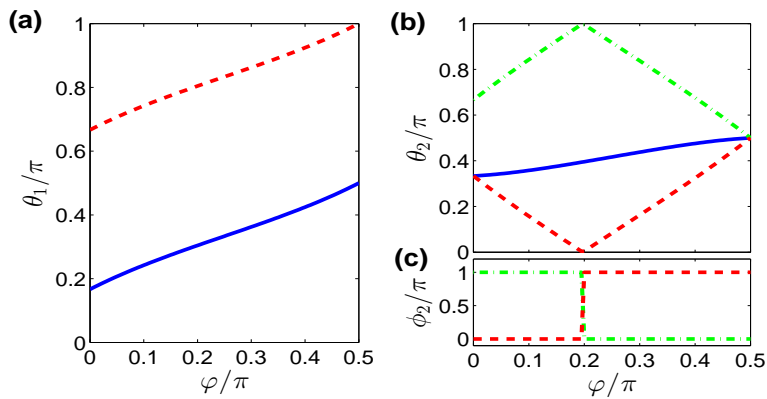}}
\caption{ Experimental values of phases in the preparation pulse $U_C(\theta_1,\phi_1)$ and the measurement pulse $U_C(\theta_2,\phi_2)$ for $\mathcal{A}=\sigma_y$ and $\mathcal{B}=\frac{\sqrt{3}}{2}\sigma_y+\frac{1}{2}\sigma_z$ with $\chi=\pi/6$. (a) $\theta_{1}$ for 729-nm laser pulse with the prepared states of $\rho_{1}$ and $\rho_{2}$ denoted by solid and dashed curves, respectively, and the phases $\phi_{1}$ being zero. (b)  $\theta_{2}$ in the measurement pulse $U_C(\theta_2,\phi_2)$ with $M_{++}$, $M_{+-}$ and $M_{-+}$ denoted by solid, dashed and dashed-dotted curves, respectively. For clarity, we omit the curves for the cases of $A_+$ and $B_+$, in which $\theta_{2}$ are fixed to be $\pi$ and $\pi/3$, respectively.  (c) $\phi_{2}$ in the measurement pulse $U_C(\theta_2,\phi_2)$ with $M_{+-}$ and $M_{-+}$ denoted by dashed and dashed-dotted curves, respectively. For clarity, we omit the curves for the cases of $M_{++}$, $A_+$ and $B_+$, in which $\phi_{2}$ is fixed to be zero. }
 \label{fig3}
\end{figure*}

The single-qubit experimental steps for optimal joint measurement is schematically illustrated in Fig.~\ref{fig2}(c). We initialise the ion in the state \mbox{$|\downarrow\rangle$} by optical pumping, and then prepare the ion to the target state $|\zeta\rangle$ (see Appendix A) which is determined by $\rho_{1,2}$ by a carrier-transition pulse $U_{C}(\theta_1,\phi_1)$. The next step is to measure the observables $A_{\pm}$, $B_{\pm}$ and $M_{\pm,\pm}$. The measurement process includes a measurement pulse $U_{C}(\theta_2,\phi_2)$ steering the state from $|\zeta\rangle$ to $|\xi\rangle$, followed by a detection. The two $U_{c}$ operations take several $\mu$s by the 729-nm laser, respectively. In general, the pulse lengths for the successive two processes are smaller than a Rabi period ($2\pi$), implying a duration less than 18 $\mu$s. Finally, the probability of finding the ion in the \mbox{$|\uparrow\rangle$} state is detected by collecting light scattered on the dipole transition and counting the emitted photons for 4 ms by the photon multiplier tube. We exemplify $\mathcal{A}=\sigma_y$ and $\mathcal{B}=\frac{\sqrt{3}}{2}\sigma_y+\frac{1}{2}\sigma_z$ to illustrate the details in the process of the optimal joint measurement. The relevant phases $\theta_{1,2}$ and $\phi_{1,2}$ for the $U_{c}$ operations are indicated in Fig.~\ref{fig3}(a-c). \\

\begin{figure*}[htbp]
\centering {\includegraphics[width=11.0 cm, height=8.5 cm]{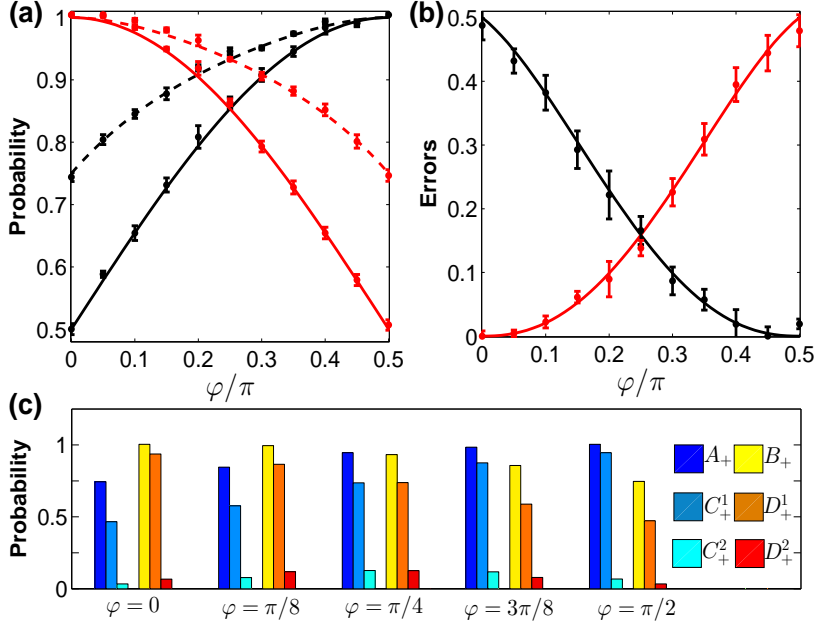}}
\caption{Experimental observation for joint measurement of $\mathcal{A}=\sigma_y$ and $\mathcal{B}=\frac{\sqrt{3}}{2}\sigma_y+\frac{1}{2}\sigma_z$ with $\chi=\pi/6$. (a) Experimental measurements of the positive operators with respect to $\varphi$. The black dashed and solid curves represent the results of theoretical prediction of $p^{A_+}_{\rho_1}$ and $p_{\rho_1}^{C_+}$, and the red dashed and solid curves denote $p^{B_+}_{\rho_2}$ and $p_{\rho_2}^{D_+}$, respectively. (b) Experimental results of the OWC errors $(\varepsilon_a,\varepsilon_b)$ corresponding to (a). The black curve ($\varepsilon_a$) and the red curve ($\varepsilon_b$) represent the results of theoretical prediction of Eq.~(\ref{eq3}). (c) Probability distributions of the positive operators for different $\varphi$.  Here $A_+$, $C_+^1=M_{++}$ and $C_+^2=M_{+-}$ are measured on $\rho_1$, and $B_+$, $D_+^1=M_{++}$ and $D_+^2=M_{-+}$ are measured on $\rho_2$. The error bars indicate standard deviation containing the statistical errors of 40,000 measurements for each data point.}
 \label{fig4}
\end{figure*}

\subsection*{3.2 Experimental observation of the qubit MUR}
We focus on the data set in Fig.~\ref{fig4}(a,b) for $\langle A_{+}\rangle$, $\langle C_+\rangle$, $\langle B_+\rangle$ and $\langle D_+\rangle$ with $\chi=\pi/6$ under the optimal approximation. With the increase of $\varphi$, one has a better approximation of $\mathcal{C}$ to $\mathcal{A}$ while the difference of $\mathcal{D}$ from $\mathcal{B}$ becomes larger, reflecting the error trade-off for these two incompatible observables. In the limit of $\mathcal{C}$ or $\mathcal{D}$ becoming a perfect approximation, the error for the other reaches its maximum required by the compatibility of $\mathcal{C},\mathcal{D}$ and given the incompatibility of $\mathcal{A},\mathcal{B}$. The experimentally observed optimal errors $(\varepsilon_a,\varepsilon_b)$, plotted in Fig.~\ref{fig4}(b), agree well with the theoretical prediction. Under the condition of $f(\bm{c},\bm{d})=2$ for optimal approximations, the maximal value of the OWC error along the optimal error curve is $\sin\chi$ (reached when the other error is zero), which is a direct measure of the incompatibility of the pair of the incompatible observables: both the error peaks and $\sin\chi$ are 0.5 in Fig.~\ref{fig4}(b). By contrast, in the case of $f(\bm{c},\bm{d})<2$, the peak values of errors will be larger than $\sin\chi$. In any case, the approximations of $\mathcal{A}$ by $\mathcal{C}$ and of $\mathcal{B}$ by $\mathcal{D}$ are still subject to the error trade-off relation of Eq. (\ref{Eq1}). To further understand the observation of the positive operators, we may check Fig.~\ref{fig4}(c) for the unified
probability distributions of the measurements.

\begin{figure*}[htpb]
\centering{\includegraphics[width=15.0 cm, height=17.0 cm]{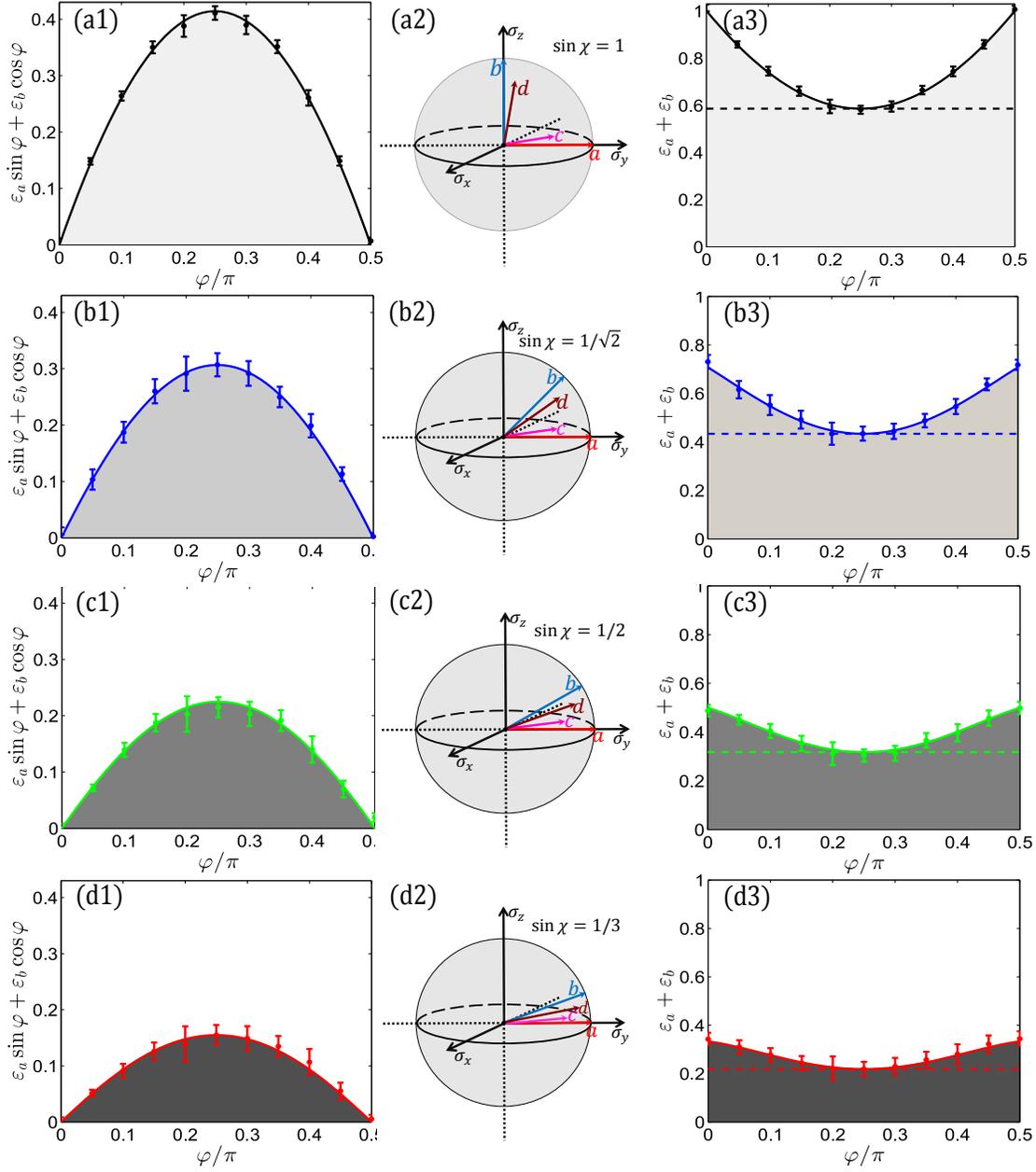}}
\caption{Experimental observation of the optimal error trade-off relation with different $\chi$. (a1-d1) Results relevant to Eq.~(\ref{Eq1}). The solid curves are for the optimal lower bounds $\sqrt{1+\sin\chi\sin 2\varphi}-1$. The shaded regions represent the forbidden areas. (a2-d2) Corresponding vectors of the observables in Bloch sphere. (a3-d3) Results relevant to Eq.~(\ref{Eq2}). The solid curves denote the analytical results of $\varepsilon_a+\varepsilon_b$, the dashed lines represent the lower bounds $\sqrt{2}(\sqrt{1+\sin\chi}-1)$ and the shaded regions represent the forbidden areas. The dots in the left-hand side and right-hand side panels are experimental values. From top to bottom, the panels represent $\sin\chi=1, \sqrt{2}/2, 1/2$ and $1/3$, respectively. Since we set $\mathcal{A}=\sigma_y$, the panels from top to bottom correspond to $\mathcal{B}=(2\sqrt{2}\sigma_y+\sigma_z)/3$, $(\sqrt{3}\sigma_y+\sigma_z)/2$, $\sqrt{2}(\sigma_y+\sigma_z)/2$ and $\sigma_z$. Each data point consists of 40,000 measurements and error bars are given by standard deviation.}
 \label{fig5}
\end{figure*}

To experimentally test the universality of the error trade-off relation, we have explored different pairs of incompatible observables and measured corresponding optimal errors as shown in Fig.~\ref{fig5}. In the left-hand side panels of Fig.~\ref{fig5}, the lower bounds examined experimentally are in good agreement with the theoretical prediction by Eq.~(\ref{Eq1}). The regions under the curves represent corresponding forbidden areas, enforced by the incompatibility quantity $\sin\chi$ in accordance with the MUR. As plotted in the right-hand side panels of Fig.~\ref{fig5}, the experimental witness of Eq.~(\ref{Eq2}) also fits well the theoretical prediction, where each lower bound reaches the minimum at $\varphi=\pi/4$, where $\mathcal{A}$ and $\mathcal{B}$ are approximated with equal errors.

\subsection*{3.3 Experimental imprecision}
In our experiment, the typical errors are from imperfection of initial-state preparation and the final-state detection,
and from heating due to the radial phonons as well as from the statistical errors. The former three
of the imperfection factors are experimentally determined errors, which can be partially corrected. In contrast,
the statistical errors are not correctable, but contained in standard deviation indicated by error bars. Moreover, the inherent decay and dephasing times of the qubit are, respectively, 1.1 s and 2 ms, whose detrimental effects are negligible during the short periods ($\sim $18$ \mu$s) of our operations.

The Rabi frequency of 729-nm laser in our system is 54(2) kHz and the occupation probability of the initial state is about 98.7(4)\%.
The detection error yields a mean deviation of 0.22(8)\%. The thermal phonons from the radial direction creates an additional dephasing effect on qubit system, yielding the depasing time of 0.24(15) ms. These imprecision can be partially calibrated by practical methods \cite {np-11-193,duan}. In contrast, the fluctuation in above three imperfections, which is not correctable, is due to instability of the laser power and the
magnetic field. The instable laser power leads to random variation of laser intensity and the small fluctuating magnetic
field shifts the resonance frequency randomly. Both the unstable factors lead to inaccuracy in initial-state preparations
and qubit operations, whose detrimental effects are assessed to be less than 2\% from the Rabi oscillation in our case and involved in the standard deviation represented by the error bars.

The statistical errors are due to quantum projection noise, a typical environment-induced noise from vacuum
fluctuation. These errors are inevitable in any quantum mechanical measurement, but can be reduced by more
measurements and/or by quantum correlation. By Monte Carlo simulation, we assess the statistical deviation for measuring the probability distribution to be 0.0025 and for each of the error pairs to be less than 0.01. These imperfections are involved in the standard deviation represented by the error bars.

\section*{4. Conclusion}

An appropriate understanding of  the uncertainty relations is essential to our exploration of new physics and precision measurements. There have been suggestions that enhanced measurement precision may be achievable by beating the so-called standard quantum limit through strategies of getting around the uncertainty relations \cite {science-306-1330}. It has also been found that the uncertainty relations can contribute to a deeper understanding of non-locality \cite {nonlocal1,nonlocal2}. Uncertainty relations have also been employed to prove the security of quantum key distribution \cite {qkey} and explore the influence of quantum memory \cite {nphys6-659}.

Our experiment provides the first evidence of confirming the MUR in  an optimal joint measurement on a pure quantum system - a single ultracold trapped-ion system. The tests performed cover a range of choices of target observables and demonstrate optimality by sampling joint measurements with error pairs distributed along the whole lower boundary of the admissible error region.

Previous theoretical proposals and experimental tests of error trade-off relations (\cite{pra67-042105}-\cite{pra85-062117}  and  \cite{natphys8-185}-\cite{prl112-020402}) are based on the EDRs that are not generally amenable to a direct comparison of the approximating and target observables---unless one restricts the class of the former to those that commute with the latter; in that case, one cannot consider the trade-off relations in question to be universal \cite{BuSte15}.
In contrast, the determination of the OWC errors in the present experiment is obtained through a direct comparison of the statistics of the observables $\mathcal{C}$ and $\mathcal{D}$ with those of $\mathcal{A}$ and $\mathcal{B}$. Our experiment therefore constitutes the first direct test of a MUR. In practice, our methods and findings, verifying the optimal joint measurements, are of potential applications in quantum information science, e.g., by providing new calibration protocols for quantum precision measurements or by leading to new ways of guaranteeing the information security in quantum cryptographic protocols.

\section*{Acknowledgments}
This work was supported by National Natural Science Foundation of China under Grant Nos. 11674360, 11404377 and 11371247, and by the Strategic Priority Research Program of the Chinese Academy of Sciences under Grant No. XDB21010100. TPX and LLY contributed equally to this work.\\

\appendix

\section{Time evolution and laser phases}
In our experiment, the process consists of two steps: one is to prepare the desired quantum state $|\zeta\rangle$ in Fig. \ref{fig2}(c), and the other is to realise the measurement operator relevant to $|\xi\rangle$ in Fig. \ref{fig2}(c).

Under the carrier transition for the first pulse, we reach the state $\rho=U_C(\theta_1,\phi_1)|\downarrow\rangle\langle\downarrow | U^{\dagger}_C(\theta_1,\phi_1)=(I+\bm{r}\cdot\bm{\sigma})/2$, where
\begin{equation}
\theta_1=2\arccos\sqrt{\frac{1-r_z}{2}}, \phi_1=\frac{\pi}{2}[1+ sign(r_y)]+\arctan\frac{r_x}{r_y},
\label{eq13}
\end{equation}
with sign$(r_y)=1$ if $r_y> 0$, sign$(r_y)=0$ if $r_y=0$ and sign$(r_y)=-1$ if $r_y<0$ under the condition of $r_{x}\neq$0; in the case of $r_{x}=0$, $\theta_1$ remains unchanged but $\phi_{1}$ turns to be zero. This operation corresponds to the step from $|\downarrow\rangle$ to $|\zeta\rangle$ in Fig. \ref{fig2}(c). After finishing this step, we obtain two states $\bm{r_1}$ and $\bm{r_{2}}$, obeying  $\bm{r_1}=(\bm{a}-\bm{c})/\Vert \bm{a}-\bm{c}\Vert$ and $\bm{r_2}=(\bm{b}-\bm{d})/\Vert \bm{b}-\bm{d}\Vert$, respectively.

The second pulse yields the measurement operator expressed as $E=(I+\bm{e}\cdot\bm{\sigma})/2=U^{\dagger}_C(\theta_2,\phi_2)|\uparrow\rangle\langle\uparrow | U_C(\theta_2,\phi_2)$, where
\begin{equation}
\theta_2=2\arccos\sqrt{\frac{1+e_z}{2}},\phi_2=\frac{\pi}{2}[1+ sign(e_y)]+\arctan\frac{e_x}{e_y}.
\label{eq14}
\end{equation}
We thus have five necessary measurement operators $E=A_+$, $B_+$, $S_+^+$, $S_{+}^-$ and $S_{-}^-$. With this operation, we reach the state $|\xi\rangle$, as plotted in Fig. \ref{fig2}(c).\\

\end{document}